\documentstyle[12pt,axodraw]{article}

\makeatletter
\def\appendix{\par\clearpage
  \setcounter{section}{0}
  \setcounter{subsection}{0}
  \@addtoreset{equation}{section}
  \def\@sectname{Appendix~}
  \def\theequation{\thesection.\arabic{equation}}
  \def\thesection{\Alph{section}}}
\makeatother

\begin{document}

\begin{titlepage}
\hskip 11cm \vbox{
\hbox{BUDKERINP/98-94}
\hbox{UNICAL-TP 98/7}
\hbox{December 1998}}
\vskip 0.3cm
\centerline{\bf THE QUARK PART OF THE NON-FORWARD BFKL KERNEL}
\centerline{\bf AND THE ``BOOTSTRAP'' FOR THE GLUON REGGEIZATION$^{~\ast}$} 
\vskip 1.0cm
\centerline{  V.S. Fadin$^{a~\dagger}$, R. Fiore$^{b~\ddagger}$ 
and A. Papa$^{b~\ddagger}$ }
\centerline{\sl $^{a}$ Budker Institute for Nuclear Physics,}
\centerline{\sl Novosibirsk State University, 630090 Novosibirsk, Russia}
\vskip 0,5cm
\centerline{\sl $^{b}$ Dipartimento di Fisica, Universit\`a della Calabria,}
\centerline{\sl Istituto Nazionale di Fisica Nucleare, Gruppo collegato di Cosenza,}
\centerline{\sl Arcavacata di Rende, I-87036 Cosenza, Italy}
\vskip 1cm

\begin{abstract}
We calculate the quark part of the kernel of the generalized non-forward
BFKL equation at non-zero momentum transfer $t$ in the next-to-leading
logarithmic approximation. Along with the quark contribution to the gluon
Regge trajectory, this part includes pieces coming from the quark-antiquark
production and from the quark contribution to the radiative corrections in
one-gluon production in the Reggeon-Reggeon collisions. The results obtained
can be used for an arbitrary representation of the colour group in the 
$t-$channel. Using the results for the adjoint representation, we demonstrate
explicitly the fulfillment of the ``bootstrap'' condition for the gluon
Reggeization in the next-to-leading logarithmic approximation in the part
concerning the quark contribution.
\end{abstract}
\vfill

\hrule
\vskip.3cm

\noindent
$^{\ast}${\it Work supported in part by the Ministero italiano
dell'Universit\`a e della Ricerca Scientifica e Tecnologica, in part
by INTAS and in part by the Russian Fund of Basic Researches.}
\vfill
$ \begin{array}{ll}
^{\dagger}\mbox{{\it email address:}} &
 \mbox{FADIN~@INP.NSK.SU}\\
\end{array}
$

$ \begin{array}{ll}
^{\ddagger}\mbox{{\it email address:}} &
  \mbox{FIORE, PAPA ~@FIS.UNICAL.IT}
\end{array}
$
\vfill
\vskip .1cm
\vfill
\end{titlepage}
\eject
\textheight 210mm \topmargin 2mm \baselineskip=24pt

\newpage

\section{Introduction}

The BFKL equation~\cite{FKL} became very popular in the last years due to
the experimental results on deep inelastic scattering obtained at HERA~\cite
{H1}. These results show that the power of the growth of the cross section
of the photon-proton interaction with the energy for a ``hard'' photon (the
``hardness'' is supplied either by the photon virtuality or by the masses of
the quarks into which the photon is converted) is larger than the
corresponding power for hadron processes. The idea arose that the rapid
increase of the cross section of the ``hard'' photon interactions is the
manifestation of the BFKL dynamics.

The BFKL equation was derived in the leading logarithmic approximation (LLA)
of the QCD perturbation theory, which means summation of all terms of the
type $[\alpha _{s}\ln s]^{n}$; $\alpha _{s}$ is the QCD coupling constant
and $s$ is the square of the c.m.s. energy. Unfortunately, in this
approximation neither the scale of $s$ nor the argument of the running
coupling constant $\alpha _{s}$ are fixed. So, in order to do accurate
theoretical predictions, we have to know the radiative corrections to the
LLA. The program of the calculation of the radiative corrections was
formulated in Ref.~\cite{FaLi89} and fulfilled in Refs.~\cite{FL93}-\cite
{FFFK}. Recently, the calculation of the radiative corrections to the kernel
of the BFKL equation was completed and the equation in the next-to-leading
logarithmic approximation (NLLA) was obtained~\cite{FaLi98,CC}. The corrections
appear to be large and caused a series of papers~\cite{Blue} devoted to the
problem how to deal with them and what they mean.

The BFKL equation is a particular case (for forward scattering, i.e. $
t=0 $ and vacuum quantum numbers in the $t-$channel) of the equation for the 
$t-$channel partial waves of the elastic amplitudes~\cite{FKL}.
Independently from the value of $t,$ we have in general a mixture of various
irreducible representations ${\cal R}$ of the colour group in the $t-$channel. 
The most interesting representations are the colour singlet
(Pomeron channel) and the antisymmetric colour octet (gluon channel). For
brevity, we use the term ``BFKL equation'' for the general case as well,
adding the word ``non-forward'' when it is necessary to distinguish the
general case from the particular ``forward'' case.

It is very important to find the corrections to the kernel of the non-forward BFKL,
for the gluon channel as well as for the
Pomeron channel. In the case of the Pomeron channel the equation can be
applied directly for the description of experimental data. The importance of
the correction in the gluon channel is determined by a remarkable property
of QCD, the gluon Reggeization. We remind that the derivation of the BFKL
equation was based~\cite{FKL} on this property. In fact, this equation is
the equation for the Green function of two Reggeized gluons. In the colour
singlet state these Reggeized gluons create the Pomeron. The
self-consistency requires that in the colour octet case the two Reggeized
gluons reproduce the Reggeized gluon itself (``bootstrap'' condition). The
above statements are valid in the NLLA as well as in LLA. The ``bootstrap''
equations in the NLLA were recently derived~\cite{FF98}. Since the BFKL
equation is very important for the theory of Regge processes at high energy $
\sqrt{s}$ in perturbative QCD, these equations must be checked. 
Along with the stringent test of the gluon Reggeization, this check has another
important meaning. The calculations of the radiative corrections to the kernel
are very complicated. Therefore, they should be carefully verified. Up
to now, only a small part of the calculations was independently 
performed~\cite{CCH} or checked~\cite{Blu}. The bootstrap equations
contain all the values appearing in the calculations of the NLLA kernel,
so that they provide a global test of the calculations.
Beside this, the colour octet state of two Reggeized gluons is necessary for the
description of colourless compound states of more than two gluons, in
particular, for the Odderon.

In this paper we consider the non-forward BFKL equation, calculate the quark
contribution to the kernel of this equation and demonstrate explicitly the
fulfillment of the ``bootstrap'' conditions in the NLLA in the part
concerning the quark-antiquark contribution.

In the next Section we present the general form of the quark contribution to
the kernel. In Section 3 we give the explicit form of the quark piece of the
gluon trajectory and derive the quark part of the contribution to the kernel
from the one-gluon production. In Section 4 we consider the quark-antiquark
production in collisions of two Reggeized gluons. In Section 5 we obtain the
contribution of this process to the kernel. In Section 6 we demonstrate the
fulfillment of the ``bootstrap'' condition for the trajectory. The results
obtained are discussed  in Section 7.

\section{Definitions and basic equations}

As usual, in an analysis of processes at high energy particle collisions, we
use the Sudakov decomposition for particle momenta. Admitting that the initial
particles $A$ and $B$ have non-zero masses, we introduce the light-cone
vectors $p_{+}$ and $p_{-}$ in terms of which the momenta of the initial particles
are $p_A=p_{+}+(m_A^2/s)$ $p_{-}$ and $p_B=p_{-}+(m_B^2/s)$ $p_{+}$
respectively, with $s=2(p_{+}p_{-})$. In the NLLA, as well as in the LLA, the
elastic scattering amplitudes with momentum transfer $q$ $\simeq q_{\perp }$
are expressed in terms of the impact factors $\Phi $ of the scattering
particles and of the Green function $G$ for the scattering of Reggeized
gluons~\cite{FF98} (see Fig.~1). The Mellin transform of the Green function
for Reggeized gluons with initial momenta in the $s-$channel $q_1\simeq
\beta p_{+}+q_{1\perp }$ and $-q_2\simeq \alpha p_{-}-q_{2\perp },$
momentum transfer $q$ $\simeq q_{\perp }$ and irreducible representation $
{\cal R}$ of the colour group in the $t-$channel, obeys the equation~\cite
{FF98} 
\[
\omega G_\omega ^{\left( {\cal R}\right) }\left( \vec q_1,\vec q_2;\vec
q\right) =
\]
\begin{equation}
\vec q_1^{\:2}\left( \vec q_1-\vec q\,\right) ^2\delta ^{\left( D-2\right)
}\left( \vec q_1-\vec q_2\right) +\int \frac{d^{D-2}r}{{\vec r}
_{}^{\,2}\left( {\vec r}-\vec q\,\right) ^2}{\cal K}^{\left( {\cal R}
\right) }\left( \vec q_1,{\vec r};\vec q\right) G_\omega ^{\left( {\cal R}
\right) }\left( {\vec r},\vec q_2;\vec q\,\right) ~.  \label{z1}
\end{equation}

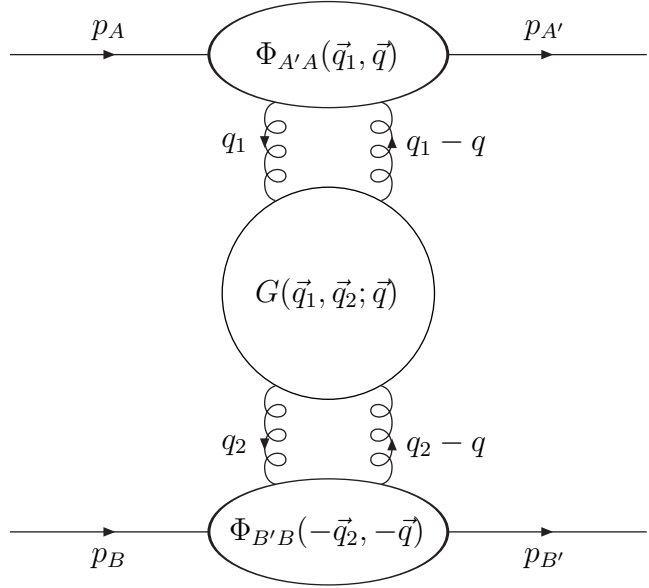
\begin{figure}[tb]	
\begin{center}
\begin{picture}(240,200)(0,0)

\ArrowLine(0,190)(75,190)
\ArrowLine(165,190)(240,190)
\Text(37.5,200)[]{$p_A$}
\Text(202.5,200)[]{$p_{A'}$}
\Text(120,190)[]{$\Phi_{A'A}(\vec q_1,\vec q)$}
\Oval(120,190)(20,45)(0)

\Gluon(100,135)(100,172){4}{3}
\Gluon(140,172)(140,135){4}{3}

\Gluon(100,28)(100,65){4}{3}
\Gluon(140,65)(140,28){4}{3}

\ArrowLine(96,158)(96,156)
\ArrowLine(144,156)(144,158)

\ArrowLine(96,44)(96,42)
\ArrowLine(144,42)(144,44)

\Text(85,157)[]{$q_1$}
\Text(165,157)[]{$q_1-q$}

\Text(85,43)[]{$q_2$}
\Text(165,43)[]{$q_2-q$}

\GCirc(120,100){40}{1}
\Text(120,100)[]{$G(\vec q_1,\vec q_2;\vec q)$}

\ArrowLine(0,10)(75,10)
\ArrowLine(165,10)(240,10)\Text(37.5,0)[]{$p_B$}
\Text(202.5,0)[]{$p_{B'}$}
\Text(120,10)[]{$\Phi_{B'B}(-\vec q_2,-\vec q)$}
\Oval(120,10)(20,45)(0)

\end{picture}
\end{center}
\caption[]{Diagrammatic representation of the elastic scattering amplitude
$A + B \rightarrow A' + B'$.} 
\end{figure}	

\noindent 
Here $\vec q_1$ and $-\vec q_2$ are the transverse momenta of the colliding gluons
in the $s-$channel, $\vec q$ is the momentum transfer and $D=4+2\epsilon $
is the space-time dimension, taken different from $4$ to regularize the
infrared divergences. Let us note that we use a normalization which is
different from the one used for the forward case~\cite{FaLi98}. The
non-forward kernel, analogously to the forward one, is given as the
sum of the ``virtual'' part, defined by the gluon trajectory $j(t)=1+\omega
\left( t\right) $, and the ``real'' part ${\cal K}_r^{\left( {\cal R}
\right) }$, related to the real particle production in Reggeon-Reggeon
collisions: 
\[
{\cal K}^{\left( {\cal R}\right) }\left( \vec q_1,\vec q_2;\vec
q\,\right) =
\]
\begin{equation}
\left[ \omega \left( -\vec q_1^{\:2}\right) +\omega \left( -\left( \vec
q_1-\vec q\,\right) ^2\right) \right] \vec q_1^{\:2}\left( \vec q_1-\vec
q\,\right) ^2\delta ^{\left( D-2\right) }\left( \vec q_1-\vec q_2\right) + 
{\cal K}_r^{\left( {\cal R}\right) }\left( \vec q_1,\vec q_2;\vec
q\right) ~.  \label{z2}
\end{equation}
The gluon trajectory is known~\cite{F} in the NLLA. The ``real'' part for
the non-forward case is known in the LLA only~\cite{FKL}: 
\begin{equation}
{\cal K}_r^{\left( {\cal R}\right) B}\left( \vec q_1,\vec q_2;\vec
q\,\right) =\frac{g^2c_R}{(2\pi )^{D-1}}\left( \frac{\vec q_1^{\:2}{(\vec
q_2-\vec q \,)}^2+\vec q_2^{\:2}{(\vec q_1-\vec q \,)}^2}{(\vec q_1- \vec
q_2)^2}-\vec q^{\:2}\right) ~,  \label{z3}
\end{equation}
where the superscript $B$ means the LLA (Born) approximation and the
coefficients $c_R$ for the singlet $(R=1)$ and octet $(R=8)$ cases are 
\begin{equation}
c_1=N,\quad c_8=\frac N2 ~,  \label{z4}
\end{equation}
$N$ being the number of colours ($N=3$ in QCD).
In Eq.~(\ref{z3}) and below $g$ is the bare coupling constant, connected
with the renormalized coupling $g_\mu $ in the ${\overline{MS}}$ scheme by
the relation 
\begin{equation}
g=g_\mu \mu ^{-\mbox{\normalsize
$\epsilon$}}\left[ 1+\left( \frac{11}3-\frac 23\frac{n_f}N\right) \frac{\bar
g_\mu ^2}{2\epsilon }\right] ~,  \label{z4a}
\end{equation}
where 
\begin{equation}
\bar g_\mu ^2=\frac{g_\mu ^2N\Gamma (1-\epsilon )}{(4\pi )^{2+{\epsilon }}}~.
\label{z4b}
\end{equation}
Let us stress that in this paper we will systematically use the perturbative
expansion in terms of the bare coupling $g$.

In the NLLA the ``real'' part of the kernel can be presented as~\cite{FF98} 
\[
{\cal K}_{r}^{\left( {\cal R}\right) }\left( \vec{q}_{1},\vec{q}_{2};
\vec{q}\,\right) =\int \frac{ds_{_{RR}}}{\left( 2\pi \right) ^{D}}{\cal I}
\,m{\cal A}_{RR}^{\left( {\cal R}\right) }\left(
q_{1},  q_{2}; \vec q\,\right) \theta \left( s_{_{\Lambda }}-s_{_{RR}}
\right) 
\]
\begin{equation}
-\frac{1}{2}\int \frac{d^{D-2}r}{{\vec{r}}^{\,2}\left( {\vec{r}}-\vec{q}
\,\right) ^{2}}{\cal K}_{r}^{\left( {\cal R}\right) B}\left( \vec{q}
_{1},{\vec{r}};\vec{q}\right) {\cal K}_{r}^{({\cal R}){\scriptsize {B}}
}\left( {\vec{r}},\,\vec{q}_{2};\,\vec{q}\,\right) \,\ln\left( \frac{
s_{_{\Lambda }}^{2}}{({\vec{r}}-\vec{q}_{1})^{2}({\vec{r}}-\vec{q}_{2})^{2}}
\right) ~.  \label{z5}
\end{equation}
Here ${\cal A}_{RR}^{\left( {\cal R}\right) }\left(
q_{1}, q_{2}; \vec q\right) $ is the scattering amplitude of the 
Reggeons with initial momenta $q_{1}=\beta p_{+}+q_{1\perp }\ $ and $\ -q_{2}=\alpha
p_{-}-q_{2\perp }$ and momentum transfer $q=q_{\perp }\,$, for the
representation ${\cal R}$ of the colour group in the $t-$channel, $
s_{_{RR}}=\left( q_{1}-q_{2}\right) ^{2}$ is the squared invariant mass of
the Reggeons. The $s_{_{RR}}-$channel imaginary part ${\cal I}m
{\cal A}_{RR}^{\left( {\cal R}\right) }\left( \vec q_{1},\vec q_{2};\vec q
\right) $ is expressed in terms of the effective vertices for the production
of particles in the Reggeon-Reggeon collisions~\cite{FF98}. The second term
in the r.h.s. of Eq.~(\ref{z5}) serves for the subtraction of the
contribution of the large $s_{_{RR}}$ region in the first term,
in order to avoid a double counting of this region in the BFKL equation. The
intermediate parameter $s_{_{\Lambda }}$ in Eq.~(\ref{z5}) must be taken
tending to infinity. At large $s_{_{RR}}$ only the contribution of the
two-gluon production does survive in the first integral, so that the
dependence on $s_{_{\Lambda }}$ disappears in Eq.~(\ref{z5}) because of the
factorization property of the two-gluon production vertex~\cite{FF98}.
Since we are interested here in the quark contribution only, we can omit the
subtraction term and perform the integration over $s_{_{RR}}$ up to infinity.

The imaginary part of the Reggeon-Reggeon scattering amplitudes entering
Eq.~(\ref{z5}) can be expressed in terms of the production vertices, with
the help of the operators $\hat {{\cal P}}_{{\cal R}}$ for the
projection of two-gluon colour states in the $t-$channel on the irreducible
representations ${\cal R}$ of the colour group. We have~\cite{FF98} 
\begin{equation}
{\cal I}m{\cal A}_{RR}^{\left( {\cal R}\right) }\left(
q_1,q_2;\vec q\right) =\frac{\langle c_1c_1^{\prime }|\hat {{\cal P}}_{
{\cal R}}|c_2c_2^{\prime }\rangle }{2n_{{\cal R}}}\sum_{\left\{
f\right\} }\int \gamma _{c_1c_2}^{\left\{ f\right\} }\left( q_1,q_2\right)
\left( \gamma _{c_1^{\prime }c_2^{\prime }}^{\left\{ f\right\} }\left(
q_1^{\prime },q_2^{\prime }\right) \right) ^{*}d\rho _f\ ~.  \label{z6}
\end{equation}
Here and below $q_i^{\prime }=q_i-q, \: i=1,2$; $n_{{\cal R}}$ is the
number of independent states in the representation ${\cal R}$, $\gamma
_{c_1c_2}^{\left\{ f\right\} }\left( q_1,q_2\right) $ is the effective
vertex for the production of particles $\left\{ f\right\} $ in collisions of 
Reggeons with momenta $q_1$, $-q_2$ and colour indices $c_1$, $c_2$ respectively, 
$d\rho _f$ is their phase space element, 
\begin{equation}
d\rho _f=\left( 2\pi \right) ^D\delta ^{\left( D\right) }(q_1-q_2- 
\mathop{\textstyle\sum}_{\left\{ f\right\} }l_f)\prod_{\left\{ f\right\} } 
\frac{d^{D-1}l_f}{\left( 2\pi \right) ^{D-1}2\epsilon _f}\ ~.  \label{z7}
\end{equation}
The sum over $\left\{ f\right\} $ in Eq.~(\ref{z6}) is performed over all
the contributing particles $\left\{ f\right\} $ and over all their discreet
quantum numbers. In the LLA only the one-gluon production does contribute and
Eq.~(\ref{z5}) gives for the kernel its LLA\ value (\ref{z3}); in the NLLA
the contributing states include also the two-gluon and the quark-antiquark
states. The normalization of the corresponding vertices is defined in 
Ref.~\cite{FF98}.

The most interesting representations ${\cal R}$ are the colour singlet
(Pomeron channel) and the antisymmetric colour octet (gluon channel). We
have for the singlet case 
\begin{equation}
\langle c_1c_1^{\prime }|\hat {{\cal P}}_1|c_2c_2^{\prime }\rangle =\frac{
\delta _{c_1c_1^{\prime }}\delta _{c_2c_2^{\prime }}}{N^2-1}\ ,\quad n_1=1\ ~,
\label{z8}
\end{equation}
and for the octet case 
\begin{equation}
\langle c_1c_1^{\prime }|\hat {{\cal P}}_8|c_2c_2^{\prime }\rangle =\frac{
f_{c_1c_1^{\prime }c}f_{c_2c_2^{\prime }c}}N\ ,\ \ \ \ n_8=N^2-1 ~,
\label{z9}
\end{equation}
where $f_{abc}$ are the structure constants of the colour group.

\section{Quark part of the kernel from the gluon trajectory and real gluon
production}

The gluon trajectory is known~\cite{F} in the NLLA. The quark contribution $
\omega _Q(t)$ to the trajectory appears at the two-loop level only. For the
case of $n_f$ massless quark flavours it can be written as 
\begin{equation}
\omega _Q^{(2)}(t)=\frac{g^2t}{(2\pi )^{D-1}}\int \frac{d^{(D-2)}q_1}{\vec
q_1^{\:2}(\vec q_1-\vec q \,)^{\:2}}\left[ F_Q(\vec q \,)-2F_Q(\vec q_1)
\right] ~,  \label{z10}
\end{equation}
where $t=q^2=-{\vec q}^{\:2}$ and 
\begin{equation}
F_Q(\vec q \,)=\frac{2g^2Nn_f\Gamma \left( 2-\frac D2\right) \Gamma ^2\left(
\frac D2\right) }{(4\pi )^{\frac D2}\Gamma \left( D\right) }\left( \vec
q^{\:2}\right) ^{(\frac D2-2)}~.  \label{z11}
\end{equation}
The quark contribution to the Reggeon-Reggeon-gluon (RRG) vertex was
calculated in Ref.~\cite{FFQ}. We remind that beyond the LLA the vertex has a
complicated analytical structure~\cite{Bart, FL93}, but in the NLLA only the
real parts of the production amplitudes do contribute, because only these
parts interfere with the LLA amplitudes, which are real. Neglecting the
imaginary parts, the quark contribution to the RRG vertex becomes 
\[
\gamma _{c_1c_2}^{G(Q)}(q_1,q_2)=\varepsilon _G^dT_{c_1c_2}^d\frac{
g^3n_f\Gamma \left( 2-\frac D2\right) \Gamma ^2\left( \frac D2-1\right) }{
(4\pi )^{\frac D2}\Gamma \left( D\right) }e_G^{*\mu }\left\{ 2C_\mu
(q_2,q_1)[f_1^{(Q)}+f_2^{(Q)}]\right.
\]
\begin{equation}
\left. +\left(\frac{p_A}{(k p_A)}-\frac{p_B}{(k p_B)}\right)_\mu
[f_3^{(Q)}-(2{\vec k} _{}^2-\vec q_1^{\:2}-\vec q_2^{\:2})f_2^{(Q)}]\right\}
~.  \label{z12}
\end{equation}
Here $\varepsilon _G^d$ and $e_G^{*\mu }$ are the polarization vectors of
the produced gluon in the colour and Minkowskii spaces respectively, 
$T^d$ is the colour group generator in the adjoint representation,
$k=q_1-q_2$ is the gluon momentum, 
\begin{equation}
C(q_2,q_1)=-q_{1\perp }-q_{2\perp }+(q_1-q_{1\perp })\left( 1-\frac{2\vec
q_1^{\:2}}{{\vec k}_{}^2}\right) +(q_2-q_{2\perp })\left( 1-\frac{2\vec
q_2^{\:2}}{{\vec k}_{}^2}\right)  \label{z12a}
\end{equation}
and
\[
f_{1}^{(Q)}=\frac{(\vec{q}_{1}^{\:2}+\vec{q}_{2}^{\:2})}{(\vec{q}_{1}^{\:2}-
\vec{q}_{2}^{\:2})}\left( \frac{D}{2}-1\right) ^{2}\phi _{0},\quad \quad
f_{2}^{(Q)}=\frac{{\vec{k}}_{{}}^{2}}{(\vec{q}_{1}^{\:2}-\vec{q}_{2}^{\:2})^{3}
}\left[ \left( \frac{D}{2}-2\right) \phi _{2}-\vec{q}_{1}^{\:2}\vec{q}
_{2}^{\:2}\frac{D}{2}\phi _{0}\right] ~,
\]
\begin{equation}
f_{3}^{(Q)}=\frac{1}{(\vec{q}_{1}^{\:2}-\vec{q}_{2}^{\:2})^{{}}}\left[ \vec{q}
_{1}^{\:2}\vec{q}_{2}^{\:2}(D-2)^{2}\phi _{0}+\left( \frac{D}{2}-2\right) {
\vec{k}}_{{}}^{2}\phi _{1}\right] ~,  \label{z13}
\end{equation}
where 
\begin{equation}
\phi _{n}=\left[ (\vec{q}_{1}^{\:2})^{n+\epsilon }-(\vec{q}
_{2}^{\:2})^{n+\epsilon }\right] ~.  \label{z14}
\end{equation}
Using the expressions (\ref{z12})-(\ref{z13}), with the help of 
Eqs.~(\ref{z6})-(\ref{z9}), we obtain from Eq.~(\ref{z5})
\[
\ {\cal K}_{RRG}^{\left( {\cal R}\right) Q}\left( \vec{q}_{1},\vec{q}
_{2};\vec{q}\,\right) =c_{R}\ \frac{g^{4}n_{f}}{(2\pi )^{D-1}}\ \frac{\Gamma
(-\epsilon )}{(4\pi )^{2+\epsilon }}\ \ \frac{[\Gamma (1+\epsilon )]^{2}}{
\Gamma (4+2\epsilon )}\ 
\]
\[
\times \left\{ \ \left[ {\vec{k}}_{{}}^{2}\left( 2{\vec{k}}_{{}}^{2}-\vec{q}
_{1}^{\:2}-\vec{q}_{2}^{\:2}-2\vec{q}_{1}^{\:\prime \:2}-2\vec{q}_{2}^{\:\prime
\:2}+2\vec{q}^{\:2}\right) +(\vec{q}_{1}^{\:2}-\vec{q}_{2}^{\:2})(\vec{q}
_{1}^{\:\prime \:2}-\vec{q}_{2}^{\:\prime \:2})\right] \right. 
\]
\[
\left. \times \frac{\left[ 2(1+\epsilon )\vec{q}_{1}^{\:2}\vec{q}
_{2}^{\:2}\phi _{0}-\epsilon (\vec{q}_{1}^{\:2}+\vec{q}_{2}^{\:2})\phi _{1}
\right] }{(\vec{q}_{1}^{\:2}-\vec{q}_{2}^{\:2})^{3}}+\frac{({\vec{k}}_{{}}^{2}-
\vec{q}_{1}^{\:\prime \:2}-\vec{q}_{2}^{\:\prime \:2})}{(\vec{q}_{1}^{\:2}-\vec{q}
_{2}^{\:2})}\epsilon \phi _{1}\right. 
\]
\[
\ \left. +2(1+\epsilon )^{2}\left( \frac{\vec{q}_{1}^{\:2}\vec{q}
_{2}^{\:\prime \:2}-\vec{q}_{2}^{\:2}\vec{q}_{1}^{\:\prime \:2}}{{\vec{k}}
_{{}}^{2}}+\frac{2\vec{q}_{1}^{\:2}\vec{q}_{2}^{\:2}-\vec{q}^{\:2}(\vec{q}
_{1}^{\:2}+\vec{q}_{2}^{\:2})}{(\vec{q}_{1}^{\:2}-\vec{q}_{2}^{\:2})}\right)
\phi _{0}\right. 
\]
\begin{equation}
\left. \ +\;\;(\vec{q}_{1}\longleftrightarrow \vec{q}_{1}^{\:\prime },\;\;
\vec{q}_{2}\longleftrightarrow \vec{q}_{2}^{\:\prime })\right\} ~.  \label{z15}
\end{equation}
In the physical limit $\epsilon \rightarrow 0$ we have 
\[
\ {\cal K}_{RRG}^{\left( {\cal R}\right) Q}\left( \vec{q}_{1},\vec{q}
_{2};\vec{q}\,\right) =-c_{R}\ \frac{g^{4}n_{f}}{24(2\pi )^{5}}\ 
\]
\[
\times \left\{ \ \left[ {\vec{k}}_{{}}^{2}\left( 2{\vec{k}}_{{}}^{2}-\vec{q}
_{1}^{\:2}-\vec{q}_{2}^{\:2}-2\vec{q}_{1}^{\:\prime \:2}-2\vec{q}_{2}^{\:\prime
\:2}+2\vec{q}^{\:2}\right) +(\vec{q}_{1}^{\:2}-\vec{q}_{2}^{\:2})(\vec{q}
_{1}^{\:\prime \:2}-\vec{q}_{2}^{\:\prime \:2})\right] \right. 
\]
\[
\left. \times \frac{\left[ 2\vec{q}_{1}^{\:2}\vec{q}_{2}^{\:2}\ln \left( \vec{q
}_{1}^{\:2}/\vec{q}_{2}^{\:2}\right) -(\vec{q}_{1}^{\:4}-\vec{q}_{2}^{\:4})
\right] }{(\vec{q}_{1}^{\:2}-\vec{q}_{2}^{\:2})^{3}}+({\vec{k}}_{{}}^{2}-\vec{q
}_{1}^{\:\prime \:2}-\vec{q}_{2}^{\:\prime \:2})\right. 
\]
\[
\ \left. +2\left( \frac{\vec{q}_{1}^{\:2}\vec{q}_{2}^{\:\prime \:2}-\vec{q}
_{2}^{\:2}\vec{q}_{1}^{\:\prime \:2}}{{\vec{k}}_{{}}^{2}}+\frac{2\vec{q}
_{1}^{\:2}\vec{q}_{2}^{\:2}-\vec{q}^{\:2}(\vec{q}_{1}^{\:2}+\vec{q}_{2}^{\:2})}{(
\vec{q}_{1}^{\:2}-\vec{q}_{2}^{\:2})}\right) \ln \left( \frac{\vec{q}_{1}^{\:2}
}{\vec{q}_{2}^{\:2}}\right) \right. 
\]
\begin{equation}
\left. \ +\;\;(\vec{q}_{1}\longleftrightarrow \vec{q}_{1}^{\:\prime },\;\;
\vec{q}_{2}\longleftrightarrow \vec{q}_{2}^{\:\prime })\right\} ~.  \label{z16}
\end{equation}
A remarkable property of the kernel, which follows from the gauge
invariance of the theory, is that it vanishes when any of the vectors $\vec{q}_{i}$ 
or $\vec{q}_{i}^{\:\prime }\ $ tends to zero. One can check that
this property is fulfilled in Eqs.~(\ref{z15}) and (\ref{z16}).

\section{The quark-antiquark production in the Reggeon-Reggeon collisions}

Let us consider the production of a quark and an antiquark with momenta $l_1$ and 
$l_2$ respectively, in collisions of two Reggeons with momenta $q_1$ and $-q_2. $
We will use the Sudakov parametrization for the produced quark and antiquark
momenta $l_1$ and $l_2$ : 
\[
l_i=\beta _ip_A+\alpha _ip_B+l_{i\perp }~,~~~~~~s\alpha _i\beta
_i=-l_{i\perp }^2~=\vec l_i{}_{}^{\:2},~~~~~~i=1,2~,
\]
\begin{equation}
\beta _1+\beta _2=\beta ~,~~~~~~\alpha _1+\alpha _2=\alpha \ ~,\qquad
l_{1\perp }+l_{2\perp }=q_{1\perp }-q_{2\perp }\quad ,~  \label{z17}
\end{equation}
and the denotation 
\begin{equation}
k=l_1+l_2=q_1-q_2~,\quad s_{_{RR}}=k^2~.  \label{z18}
\end{equation}
For the effective vertex of the quark-antiquark production in the
Reggeon-Reggeon collision we have~\cite{FFFK}
\begin{equation}
\gamma _{c_1c_2}^{Q{\overline{Q}}}(q_1,q_2)=\frac 12g^2\bar u\left(
l_1\right) \left[ t^{c_1}t^{c_2}b(l_1,l_2)\,-t^{c_2}\,t^{c_1}\overline{
b(l_2,l_1)}\right] v(l_2)~,  \label{z19}
\end{equation}
where $t^c$ are the colour group generators in the fundamental
representation. The expressions for $b(l_1,l_2)$ and $\overline{b(l_2,l_1)}$
can be presented in the following way: 
\begin{equation}
b(l_1,l_2)=\frac{4\mbox{${\not{\hbox{\kern-2.0pt$p$}}}$}_A 
\mbox{${\not{\hbox{\kern-2.0pt$Q$}}}$}_1 
\mbox{${\not{\hbox{\kern-2.0pt$p$}}}$}_B}{s~\tilde t_1}-\frac 1{{k}^2} 
\mbox{${\not{\hbox{\kern-2.0pt$\Gamma$}}}$}  \label{z20}
\end{equation}
and 
\begin{equation}
\overline{b(l_2,l_1)}=\frac{4\mbox{${\not{\hbox{\kern-2.0pt$p$}}}$}_B 
\mbox{${\not{\hbox{\kern-2.0pt$Q$}}}$}_2 
\mbox{${\not{\hbox{\kern-2.0pt$p$}}}$}_A}{s~\tilde t_2}-\frac 1{{k}^2} 
\mbox{${\not{\hbox{\kern-2.0pt$\Gamma$}}}$}~,  \label{z21}
\end{equation}
where 
\[
~\tilde t_1=(q_1-l_1)^2~,~~~~~~\tilde t_2=(q_1-l_2)^2~,
\]
\[
Q_1=q_{1\perp }-l_{1\perp }~,~~~~~~Q_2=q_{1\perp }-l_{2\perp }~,
\]
\begin{equation}
\Gamma =2\left[ (q_1+q_2)_{\perp }-\beta p_A\left( 1-2\frac{\vec q_1^{\:2}}{
s\alpha \beta }\right) +\alpha p_B\left( 1-2\frac{\vec q_2^{\:2}}{s\alpha
\beta }\right) \right] ~.  \label{z22}
\end{equation}

According to Eqs.~(\ref{z5}) and (\ref{z6}), the quark-antiquark
contribution to the BFKL kernel can be presented in the form 
\begin{equation}
{\Large {\cal K}_{RRQ{\overline{Q}}}^{\left( {\cal R}\right) }\left(
\vec q_1,\vec q_2;\vec q\,\right) }=\frac{\langle c_1c_1^{\prime }|\hat {
{\cal {P} }}_{{\cal R}}|c_2c_2^{\prime }\rangle }{2n_{{\cal R}}}
\frac{1}{(2\pi )^D}\sum_{Q{\ \overline{Q}}}\int \gamma _{c_1c_2}^{Q{\overline{Q}}
}\left( q_1,q_2\right) \left( \gamma _{c_1^{\prime }c_2^{\prime }}^{Q{
\overline{Q}}}\left( q_1^{\prime },q_2^{\prime }\right) \right)
^{*}ds_{_{RR}}d\rho _f\ ~,  \label{z23}
\end{equation}
where the sum is taken over spin, colour and flavour states of the produced
pair, $q_i^{\prime }=q_i^{}-q,\ \ s_{_{RR}}=(q_1-q_2)^2$ is the squared
invariant mass of the two Reggeons and the element $d\rho _f$ of the phase
space is given by Eq.~(\ref{z7}). The $\theta$-function in Eq.~(\ref{z5}) is
omitted since the integral over $s_{_{RR}}$ is convergent at infinity. From
the representation (\ref{z19}) we obtain 
\begin{equation}
\frac{\langle c_1c_1^{\prime }|\hat {{\cal P}}_{{\cal R}
}|c_2c_2^{\prime }\rangle }{2n_{{\cal R}}}\sum_{Q{\overline{Q}}}\gamma
_{c_1c_2}^{Q{\overline{ Q}}}\left( q_1,q_2\right) \left( \gamma
_{c_1^{\prime }c_2^{\prime }}^{Q{\ \overline{Q}}}\left( q_1^{\prime
},q_2^{\prime }\right) \right) ^{*}=\frac{ g^4n_f}{32N}[a_R^{}A+b_R^{}B+(l_1
\leftrightarrow l_2)]~.  \label{z24}
\end{equation}
Here $n_f$ is the number of light quark flavours, the coefficients $a_R$ and 
$b_R$ for the interesting cases of singlet and octet representations are 
\begin{equation}
a_0=N^2-1 ~,\quad b_0^{}=1~;\quad a_8^{}=\frac{N^2}2 ~,\quad b_8^{}=0  \label{z25}
\end{equation}
and \quad 
\[
A=tr\left( \mbox{${\not{\hbox{\kern-2.0pt$l$}}}$}_1b(l_1,l_2) 
\mbox{${\not{\hbox{\kern-2.0pt$l$}}}$}_2\overline{b^{\prime }(l_1,l_2)}
\right) ~,
\]
\begin{equation}
B=tr\left( \mbox{${\not{\hbox{\kern-2.0pt$l$}}}$}_1b(l_1,l_2) 
\mbox{${\not{\hbox{\kern-2.0pt$l$}}}$}_2b^{\prime }(l_2,l_1)\right) ~,
\label{z26}
\end{equation}
where $b^{\prime }(l_1,l_2)$ is obtained from $b(l_1,l_2)$ by the
substitution $q_{1,2}\rightarrow q_{1,2}^{\prime }\equiv q_{1,2}-q$~. In the
following, for reasons evident from Eq.~(\ref{z25}), we will call $A$ and 
$(B-A)$ ``non-Abelian'' and ``Abelian'' parts, respectively. 

The calculation of the traces gives 
\[
A=32\frac{s\alpha _1\beta _2}{\tilde t_1\tilde t_1^{\prime }}(\vec Q_1\vec
Q_1^{\:\prime })+8-\frac{16}{k^2}(\vec q^{\:2}-\frac{\vec q_1^{\:2}\vec
q_2^{\:\prime \:2}+\vec q_2^{\:2}\vec q_1^{\:\prime \:2}}{s\alpha \beta })
\]
\[
-\frac 8{(k^2)^2}[s\alpha \beta _1-s\alpha _1\beta +2\vec q_1^{\:2}\frac{
\alpha _1}\alpha -2\vec q_2^{\:2}\frac{\beta _1}\beta -2\vec l_1(\vec
q_1+\vec q_2)]
\]
\[
\times [s\alpha \beta _1-s\alpha _1\beta +2\vec q_1^{\:\prime \:2}\frac{
\alpha _1}\alpha -2\vec q_2^{\:\prime \:2}\frac{\beta _1}\beta -2\vec
l_1(\vec q_1^{\:\prime }+\vec q_2^{\:\prime })]
\]
\[
+\frac{16}{k^2}\left\{ \left[ \frac 1{\tilde t_1}\left( ((\vec Q_1\vec
q_1^{\:\prime })(s\alpha _1\beta _2-\vec l_1\vec l_2)+(\vec Q_1\vec
l_2)(s\alpha _1\beta +\vec l_1(\vec q_1^{\:\prime }+\vec q_2^{\:\prime
})-2\vec q_1^{\:\prime \:2}\frac{\alpha _1}\alpha ))\right. \right. \right.
\]
\begin{equation}
\left. \left. \left. +(\vec l_1\leftrightarrow \vec l_2,\alpha
_1\leftrightarrow \beta _2,\alpha _2\leftrightarrow \beta _1,\vec
q_1^{~}\leftrightarrow -\vec q_2^{~},\vec q_1^{\:\prime }\leftrightarrow
-\vec q_2^{\:\prime }) \phantom{\frac{.}{.}} \!\! \right) \right] +\left[ 
\phantom{\frac{.}{.}} \!\! \vec q_i^{~}\leftrightarrow \vec q_i^{\:\prime }
\right] \right\} ~,  \label{z27}
\end{equation}

\[
(A-B)+(\vec l_1\leftrightarrow \vec l_2)=32 \left\{ \frac{s\alpha _1\beta _2
}{\tilde t_1\tilde t_1^{\prime }}(\vec Q_1\vec Q_1^{\:\prime })\right.
\]
\begin{equation}
\left. -\frac{(\vec l_1\vec Q_1^{})(\vec l_2\vec Q_2^{\:\prime })+(\vec
l_2\vec Q_1^{})(\vec l_1\vec Q_2^{\:\prime })-(\vec l_1\vec l_2)(\vec
Q_1^{}\vec Q_2^{\:\prime })}{\tilde t_1\tilde t_2^{\prime }}\right\}
+\left\{ \vec l_1\leftrightarrow \vec l_2 \phantom{\frac{.}{.}}\!\! \right\}
~.  \label{z28}
\end{equation}
The Abelian and non-Abelian parts possess a ``nice'' behaviour at large
transverse momenta of the produced particles and at large values of their
invariant mass, that guarantees the convergence of the integral in Eq.~(\ref
{z23}) at $\vec l_i^{\,\!\;2~}\rightarrow \infty $ and $s_{_{RR}}\rightarrow
\infty $. The only region which leads to a singularity at the physical
dimension $D=4$ is the infrared region $k^2\rightarrow 0$. This singularity
is regularized by the non-zero $\epsilon =D/2-2$ . To make the discussed
behaviour explicit, one has to take into account the relations (\ref{z17})
between longitudinal and transverse variables. The functions $A$ and $B$ can
be expressed through the transverse momenta and one ratio of the
longitudinal momenta. Choosing this ratio as 
\begin{equation}
x=\frac{\beta _1}\beta ~,  \label{z29}
\end{equation}
we have 
\[
\frac{\beta _2}\beta =1-x ~,\quad \frac{\alpha _1}\alpha =\frac{(1-x)\vec
l_1^{\;2}}\Sigma ~,\quad \frac{\alpha _2}\alpha =\frac{x\vec l_2^{\;2} }
\Sigma ~,\quad
\]
\begin{equation}
s\alpha \beta =\frac \Sigma {x(1-x)}~,\quad s_{_{RR}}=k^2=\frac{\vec \Lambda^2
} {x(1-x)}~,  \label{z30}
\end{equation}
where 
\begin{equation}
\vec \Lambda =(1-x)\vec l_1-x\vec l_2,\quad \Sigma =\vec \Lambda^2 +x(1-x){
\vec k}_{}^2 ~,  \label{z31}
\end{equation}
and 
\[
\tilde t_1=-\frac{\vec l_1^{\;2}}x-\vec q_1^{\:2}+2(\vec l_1\vec
q_1^{~})~,\quad \tilde t_2=-\frac{\vec l_2^{\;2}}{1-x}-\vec q_1^{\:2}+2(\vec
l_2\vec q_1^{~})~,
\]
\begin{equation}
\tilde t_1^{\prime }=-\frac{\vec l_1^{\;2}}x-\vec q_1^{\:\prime \:2}+2(\vec
l_1\vec q_1^{\:\prime ~})~,\quad \tilde t_2^{\prime }=-\frac{\vec l_2^{\;2}}{ 
1-x}-\vec q_1^{\:\prime \:2}+2(\vec l_2\vec q_1^{\:\prime ~})~.  \label{z32}
\end{equation}
Using these relations, we obtain 
\[
A=16x(1-x)\left\{ -x(1-x)\left( 2\frac{(\vec \Lambda \vec q_1)}{\vec \Lambda
^2}+\frac{2(\vec q_1\vec l_1)-\vec q_1^{\:2}}{x\tilde t_1}\right) \left( 2 
\frac{(\vec \Lambda \vec q_1^{\:\prime ~})}{\vec \Lambda ^2}+\frac{2(\vec
q_1^{\:\prime ~}\vec l_1)-\vec q_1^{\:\prime \:2}}{x\tilde t_1^{\prime }}
\right) \right.
\]
\[
\left. -\frac{\vec q_{}^{\:2}}2\left( \frac 1{\vec \Lambda ^2}+\frac
1{x\tilde t_1}+\frac{2(\vec q_1\vec l_1)-\vec q_1^{\:2}}{x\tilde t_1\tilde
t_1^{\prime }}\right) +\frac{2(\vec \Lambda \vec k)}{\vec \Lambda ^2}\frac{
(\vec q_1\vec q_1^{\:\prime ~})-\vec q_1^{\:\prime \:2}}{\tilde t_1}+\frac{
2(\vec \Lambda \vec q_1)(\vec q_1^{\:\prime }\vec q_2^{\:\prime })}{\vec
\Lambda ^2}\left( \frac 1{\tilde t_1}-\frac 1{\tilde t_1^{\prime }}\right)
\right.
\]
\[
\left. +\frac{\vec q_1^{\:\prime \:2}}{\Sigma ^{}}\left( 4x(1-x)\frac{(\vec
\Lambda \vec q_1)}{\vec \Lambda ^2}+\frac{2(1-x)(2(\vec q_1\vec l_1)-\vec
q_1^{\:2})+\vec q_2^{\:2}}{\tilde t_1}\right) \left( 1-2x+2x(1-x)\frac{(\vec
\Lambda \vec k)}{\vec \Lambda ^2}\right) \right.
\]
\begin{equation}
\left. +x(1-x)\frac{\vec q_1^{\:2}\vec q_2^{\:\prime \:2}}{\vec \Lambda
^2\Sigma }-x(1-x)\frac{\vec q_1^{\:2}\vec q_1^{\:\prime \:2}}{\Sigma ^2}
\left( 1-2x+2x(1-x)\frac{(\vec \Lambda \vec k)}{\vec \Lambda ^2}\right)
^2\right\} +\left\{ \vec q_i^{~}\leftrightarrow \vec q_i^{\:\prime }
\phantom{\frac{.}{.}}\!\! \right\} ~,  \label{z33}
\end{equation}

\[
(A-B)+(l_{1}\leftrightarrow l_{2})=16x(1-x)\left\{ -x(1-x)\left( \frac{2(
\vec{q}_{1}\vec{l}_{1})-\vec{q}_{1}^{\:2}}{x\tilde{t}_{1}}+\frac{2(\vec{q}_{1}
\vec{l}_{2})-\vec{q}_{1}^{\:2}}{(1-x)\tilde{t}_{2}}\right) \right. 
\]
\[
\left. \times \left( \frac{2(\vec{q}_{1}^{\:\prime ~}\vec{l}_{1})-\vec{q}
_{1}^{\:\prime \:2}}{x\tilde{t}_{1}^{\prime }}+\frac{2(\vec{q}_{1}^{\:\prime ~}
\vec{l}_{2})-\vec{q}_{1}^{\:\prime \:2}}{(1-x)\tilde{t}_{2}^{\prime }}\right) +
\frac{\vec{q}_{{}}^{\:2}(2(\vec{q}_{1}\vec{l}_{1})-\vec{q}_{1}^{\:2})}{2\tilde{
t}_{1}}\left( \frac{1}{(1-x)\tilde{t}_{2}^{\prime }}-\frac{1}{x\tilde{t}
_{1}^{\prime }}\right) \right. 
\]
\[
\left. +\frac{\vec{q}_{{}}^{\:2}(2(\vec{q}_{1}^{\:\prime ~}\vec{l}_{1})-\vec{q}
_{1}^{\:\prime \:2})}{2\tilde{t}_{1}^{\prime }}\left( \frac{1}{(1-x)\tilde{t}
_{2}^{{}}}-\frac{1}{x\tilde{t}_{1}^{{}}}\right) +\frac{1}{x(1-x)\tilde{t}
_{1}^{{}}\tilde{t}_{2}^{\prime }}\left( -2(\vec{q}_{1}\vec{l}_{1})(\vec{q}
_{1}^{\:\prime }\vec{q}_{2}^{\:\prime ~})\right. \right. 
\]
\begin{equation}
\left. \left. +2(\vec{q}_{1}^{\:\prime ~}\vec{l}_{1})(\vec{q}_{1}\vec{q}
_{2})+(\vec{q}_{1}^{\:\prime \:2}-\vec{q}_{1}^{\:2})(\vec{l}_{1}\vec{k})+\vec{q}
_{1}^{\:2}\vec{q}_{2}^{\:\prime \:2}-\frac{{\vec{k}}_{{}}^{2}\vec{q}_{{}}^{\:2}}{ 
2}\right) \right\} +\left\{ l_{1}\leftrightarrow l_{2}\phantom{\frac{.}{.}} 
\!\!\right\} ~.  \label{z34}
\end{equation}

It is easy to see from Eqs.~(\ref{z33}) and (\ref{z34}) that the integrand in
Eq.~(\ref{z23}) falls down as $\left| \vec l_i\right| ^4$ when $\left| \vec
l_i\right| \rightarrow \infty $ . Taking into account that 
\begin{equation}
\int \frac{dk^2}{(2\pi )}d\rho _f=\int_0^1\frac{dx}{2x(1-x)}\int \frac{
d^{D-2}l_1}{(2\pi )^{(D-1)}} ~,  \label{z35}
\end{equation}
we see that the integration is well convergent at large $\left| \vec
l_i\right| $. As for the regions of values of $x$ close to 0 or 1 (which
correspond to large invariant masses $s_{_{RR}}$ -  see Eqs.~(\ref{z30})), the
convergence of the integration is guaranteed by the
vanishing of the functions $A$ and $B$ as $x(1-x)$, as it is evident from
Eqs.~(\ref{z33}) and (\ref{z34}). The limit $k^2\rightarrow 0$ means $\vec
\Lambda ^2\rightarrow 0$, according to Eq.~(\ref{z30}). The Abelian part is regular
in this limit, as it can be seen from Eq.~(\ref{z34}). As for the non-Abelian
part, from Eq.~(\ref{z33}) we have for its singular part 
\begin{eqnarray*}
A_{sing} &=&16x(1-x)\left\{ -8x(1-x)\left( \frac{(\vec \Lambda \vec q_1)}{ 
\vec \Lambda ^2}-\frac{\vec q_1^{\:2}(\vec \Lambda \vec k)}{\vec \Lambda ^2{ 
\vec k}_{}^2}\right) \left( \frac{(\vec \Lambda \vec q_1^{\:\prime ~})}{\vec
\Lambda ^2}-\frac{\vec q_1^{\:\prime \:2}(\vec \Lambda \vec k)}{\vec \Lambda
^2{\vec k}_{}^2}\right) \right. \\
\end{eqnarray*}
\begin{equation}
\left. -\frac{\vec q_{}^{\:2}}{\vec \Lambda ^2}+\frac{\vec q_1^{\:2}\vec
q_2^{\:\prime \:2}+\vec q_2^{\:2}\vec q_1^{\:\prime \:2}}{\vec \Lambda
^2\vec k^{\:2}}\right\} ~.  \label{z36}
\end{equation}

\section{The quark contribution to the kernel}

To obtain the contribution to the kernel ${\cal K}_{RRQ{\overline{Q}}
}^{\left( {\cal R}\right) }\left( \vec q_1,\vec q_2;\vec q\,\right) $ 
{\Large \ }from the real quark-antiquark production, we have to perform the
integration in Eq.~(\ref{z23}). Though the expressions (\ref{z33}) and (\ref
{z34}) are convenient for analyzing the behaviour of the functions $A$ and $ 
B $, they are not suitable 
for the integration with the measure given in Eq.~(\ref
{z35}). For this purpose, it is better to use the representations (\ref{z27}) 
and (\ref{z28}) which are explicitly invariant under the ``left-right''
transformation 
\begin{equation}
\vec l_1\leftrightarrow \vec l_2 ~,\quad \alpha _1\leftrightarrow \beta
_2 ~,\quad \alpha _2\leftrightarrow \beta _1 ~,\quad \vec q_1^{~}\leftrightarrow
-\vec q_2^{~}~,\quad \vec q_1^{\:\prime }\leftrightarrow -\vec q_2^{\:\prime } ~,
\label{z37}
\end{equation}
and to exploit also the integration measure in the alternative form: 
\begin{equation}
\int \frac{dk^2}{(2\pi )}d\rho _f \,=\int_0^1\frac{dy}{2y(1-y)} \int \frac{ 
d^{D-2}l_2}{(2\pi )^{(D-1)}}~,  \label{z38}
\end{equation}
where 
\begin{equation}
y=\frac{\alpha _2}\alpha ~.  \label{z39}
\end{equation}
The details of the integration are presented in Appendix I. The integration
for the non-Abelian contribution can be performed for arbitrary space-time
dimension. The result is 
\begin{eqnarray*}
\int \frac{dk^2}{(2\pi )}d\rho _fA={\Large \ \frac{16}{(4\pi )^{2+\epsilon }}
\frac{\Gamma (1-\epsilon )}\epsilon \frac{[\Gamma (1+\epsilon )]^2}{\Gamma
(4+2\epsilon )}}
\end{eqnarray*}
\[
\times \left\{ 2{\Large (1+\epsilon )}^2{\Large \left( \frac{({\vec k}
_{}^2)^\epsilon }{{\vec k}_{}^2}(\vec q_1^{\:2}\vec q_2^{\:\prime \:2}+\vec
q_2^{\:2}\vec q_1^{\:\prime \:2})+(\vec q^{\:2})^{1+\epsilon }\right) } 
\right.
\]
\[
\left. +\left[ {\vec k}_{}^2\left( 2{\vec k}_{}^2-\vec q_1^{\:2}-\vec
q_2^{\:2}-2\vec q_1^{\:\prime \:2}-2\vec q_2^{\:\prime \:2}+2\vec
q^{\:2}\right) +(\vec q_1^{\:2}-\vec q_2^{\:2})(\vec q_1^{\:\prime \:2}-\vec
q_2^{\:\prime \:2})\right] \right.
\]
\[
\left. \times \frac{\left[ 2(1+\epsilon )\vec q_1^{\:2}\vec q_2^{\:2}\phi
_0-\epsilon (\vec q_1^{\:2}+\vec q_2^{\:2})\phi _1\right] }{(\vec
q_1^{\:2}-\vec q_2^{\:2})^3}+\frac{\epsilon ({\vec k}_{}^2-\vec
q_1^{\:\prime \:2}-\vec q_2^{\:\prime \:2})-4(1+\epsilon )^2\vec q^{\:2}}{ 
(\vec q_1^{\:2}-\vec q_2^{\:2})}\phi _1\right.
\]
\begin{equation}
\left. \ +4(1+\epsilon )^2\frac{\vec q_1^{\:2}\vec q_2^{\:2}}{(\vec
q_1^{\:2}-\vec q_2^{\:2})}\phi _0+\;\;(\vec q_1\longleftrightarrow \vec
q_1^{\:\prime ~},\;\;\vec q_2\longleftrightarrow \vec q_2^{\:\prime
~})\right\}~,  \label{z40}
\end{equation}
where the functions $\phi _n$ are given in Eq.~(\ref{z14}). Considering the
physical limit $\epsilon \rightarrow 0$, we must take into account the
subsequent integration of the kernel over ${\vec k}$, which leads to
contributions $\sim \epsilon ^{-1}$ from the terms having the singularity at 
${\vec k}_{}^2=0$. Conserving all the terms giving non-zero contributions in
the limit $\epsilon \rightarrow 0$ after integration over ${\vec k}$, the result 
(\ref{z40}) in this limit reads 
\begin{eqnarray*}
\int \frac{dk^2}{(2\pi )}d\rho _fA &=&{\Large \frac 2{3(2\pi )^2}}\left\{ 
{\Large -\frac{12}{(4\pi )^\epsilon }\Gamma (-\epsilon )\frac{[\Gamma
(2+\epsilon )]^2}{\Gamma (4+2\epsilon )}\left( \frac{({\vec k}
_{}^2)^\epsilon }{{\vec k}_{}^2}(\vec q_1^{\:2}\vec q_2^{\:\prime\:2}+\vec
q_2^{\:2}\vec q_1^{\:\prime \:2})-(\vec q^{\:2})^{1+\epsilon }\right) } 
\right. ^{} \\
&&\left. +\left[ {\vec k}_{}^2\left( 2{\vec k}_{}^2-\vec q_1^{\:2}-\vec
q_2^{\:2}-2\vec q_1^{\:\prime \:2}-2\vec q_2^{\:\prime \:2}+2\vec
q^{\:2}\right) +(\vec q_1^{\:2}-\vec q_2^{\:2})(\vec q_1^{\:\prime \:2}-\vec
q_2^{\:\prime \:2})\right] \right.
\end{eqnarray*}
\[
\left. \times \frac{\left[ 2\vec q_1^{\:2}\vec q_2^{\:2}\ln \left( \vec
q_1^{\:2}/\vec q_2^{\:2}\right) -(\vec q_1^{\:4}-\vec q_2^{\:4})\right] }{ 
(\vec q_1^{\:2}-\vec q_2^{\:2})^3}+({\vec k}_{}^2-\vec q_1^{\:\prime
\:2}-\vec q_2^{\:\prime \:2})\right.
\]
\[
\ \left. +2\left( \frac{2\vec q_1^{\:2}\vec q_2^{\:2}-\vec q^{\:2}(\vec
q_1^{\:2}+\vec q_2^{\:2})}{(\vec q_1^{\:2}-\vec q_2^{\:2})}\right) \ln
\left( \frac{\vec q_1^{\:2}}{\vec q_2^{\:2}}\right) -2\vec q^{\:2}\ln \left( 
\frac{\vec q_1^{\:2}\vec q_2^{\:2}}{\vec q^{\:4}} \right) \right.
\]
\begin{equation}
\left. \ +\;\;(\vec q_1\longleftrightarrow \vec q_1^{\:\prime },\;\;\vec
q_2\longleftrightarrow \vec q_2^{\:\prime })\right\} ~.  \label{z41}
\end{equation}

The Abelian contribution is not singular at all, so we can consider it
from the beginning in the physical space-time dimension, $\epsilon =0$.
Nevertheless, this contribution has a form much more complicated than the
non-Abelian one. Evidently, the circumstance that the latter contribution is 
simpler must be related to the special role played by the gluon channel in
presence of gluon Reggeization. In
fact, the Abelian contribution was calculated many years ago~\cite{cheng} in the
framework of Quantum Electrodynamics and we can use the results obtained
there. We have   
\begin{equation}
\int \frac{dk^{2}}{(2\pi )}d\rho _{f}(A-B+l_{1}\longleftrightarrow l_{2})= 
 \frac{32}{(2\pi )^{2}}K_{1}\left(\vec{q}_{1}-\frac{\vec{q}}{2},\vec{q} 
_{2}-\frac{\vec{q}}{2}\right)~, \label{z42}
\end{equation}
with the function $K_{1}$ given by Eq.~(A39) of Ref. \cite{cheng}, where
in the r.h.s. we have to make the substitutions 
\[
\vec{q} \rightarrow \vec{q}_{1}-\frac{\vec{q}}{2}~,\quad
\vec{q}^{\:\prime }\rightarrow \vec{q}_{2}-\frac{\vec{q}}{2}~,\quad
{\vec{r}\rightarrow }\frac{\vec{q}}{2}~,\quad
\vec{Q}\rightarrow \vec{q}_{1}-\vec{q}(1-y)~,\quad
\vec{Q}^{\:\prime }\rightarrow \vec{q}_{2}-\vec{q}(1-x)~.
\]
It is worthwhile to say that Eq.~(\ref{z42}) contains a non-zero fermion
mass and, at first sight, has a logarithmic singularity when the mass
tends to zero; but the singularity is spurious because of cancellations
among various terms.    

We can now consider the quark contribution ${\cal K}_{r}^{\left( {\cal R 
}\right) Q}\left( \vec{q}_{1},\vec{q}_{2};\vec{q}\,\right) ${\Large \ }to
the ``real'' part of the non-forward kernel of the BFKL equation. It was
explained already that in the NLLA this contribution is determined by the
quark correction to the one-gluon production and by the quark-antiquark
production in the Reggeon-Reggeon collisions: 
\begin{equation}
{\cal K}_{r}^{\left( {\cal R}\right) Q}\left( \vec{q}_{1},\vec{ 
q}_{2};\vec{q}\,\right) ={\cal K}_{RRG}^{\left( {\cal R}\right)
Q}\left( \vec{q}_{1},\vec{q}_{2};\vec{q}\,\right) +{\cal K}_{RRQ\overline{ 
Q}}^{\left( {\cal R}\right) }\left( \vec{q}_{1},\vec{q}_{2};\vec{q} 
\,\right) ~.  \label{z43}
\end{equation}
The first term in the r.h.s. of this equation is given by Eqs.~(\ref{z4}) and 
(\ref{z14})-(\ref{z16}); the second, by Eqs.~(\ref{z23})-(\ref{z26}) and (\ref
{z40})-(\ref{z42}). For the octet case, as it can be seen from 
Eqs.~(\ref{z23})-(\ref{z26}), the contribution from the quark-antiquark production 
contains only the non-Abelian part, which was calculated for arbitrary space-time
dimension. Since the quark correction to the one-gluon production was also
calculated for arbitrary $D,$ the quark contribution to the
``real'' part of the kernel in the gluon channel for arbitrary $\epsilon $ turns 
out to be
\[
{\Large {\cal K}_{r}^{\left( 8\right) Q}\left( \vec{q}_{1},\vec{q}_{2}; 
\vec{q}\,\right) =g^{4}n_{f}N\frac{1}{(2\pi )^{D-1}}\frac{1}{(4\pi
)^{2+\epsilon }}\frac{\Gamma (1-\epsilon )}{\epsilon }\frac{[\Gamma
(2+\epsilon )]^{2}}{\Gamma (4+2\epsilon )}}
\]
\[
\times \left\{ \frac{({\vec{k}}_{{}}^{2})^{\epsilon }}{{\vec{k}} 
_{{}}^{2}}(\vec{q}_{1}^{\:2}\vec{q}_{2}^{\:\prime \:2}+\vec{q}_{2}^{\:2}\vec{q} 
_{1}^{\:\prime \:2})+\vec q^{\:2} \left((\vec q^{\:2})^\epsilon - (\vec{q} 
_{1}^{\:2})^\epsilon +(\vec{q}_{2}^{\:2})^\epsilon \right) \right.  
\]
\begin{equation}
\left. \ -\frac{(\vec{q}_{1}^{\:2}\vec{q}_{2}^{\:\prime \:2}-\vec{q}_{2}^{\:2} 
\vec{q}_{1}^{\:\prime \:2})}{{\vec{k}}_{{}}^{2}}\left( (\vec{q} 
_{1}^{\:2})^{\epsilon }-(\vec{q}_{2}^{\:2})^{\epsilon }\right) +\;\;(\vec{q} 
_{1}\longleftrightarrow \vec{q}_{1}^{\:\prime },\;\;\vec{q} 
_{2}\longleftrightarrow \vec{q}_{2}^{\:\prime })\right\} ~.  \label{z44}
\end{equation}
It is easy to see that the expression in the curly brackets vanishes when
any of the $\vec{q}_{i}$'s or $\vec{q}_{i}^{\:\prime }$'s tends to zero, as
it should be. In the physical limit $\epsilon \rightarrow 0,$ keeping all
the terms giving non-zero contributions in this limit after integration over 
${\vec{k}}$, we obtain 
\[
{\Large {\cal K}_{r}^{\left( 8\right) Q}\left( \vec{q}_{1},\vec{q}_{2}; 
\vec{q}\,\right) =\frac{g^{4}n_{f}N}{24(2\pi )^{5}}}
\]
\[
\times \left\{ -\frac{1}{\pi ^{\epsilon }}\frac{6}{(4\pi )^{2\epsilon }} 
\Gamma (-\epsilon )\frac{[\Gamma (2+\epsilon )]^{2}}{\Gamma (4+2\epsilon )} 
\left[ {\Large \frac{({\vec{k}}_{{}}^{2})^{\epsilon }}{{\vec{k}}_{{}}^{2}}( 
\vec{q}_{1}^{\:2}\vec{q}_{2}^{\:\prime \:2}+\vec{q}_{2}^{\:2}\vec{q} 
_{1}^{\:\prime \:2})+\vec{q}^{\:2}}\left( {\Large (\vec{q}^{\:2})^{\epsilon }-}( 
\vec{q}_{1}^{\:2})^{\epsilon }-(\vec{q}_{2}^{\:2})^{\epsilon }\right) \right]
\right. 
\]

\begin{equation}
\left. \ -\frac{(\vec{q}_{1}^{\:2}\vec{q}_{2}^{\:\prime \:2}-\vec{q}_{2}^{\:2} 
\vec{q}_{1}^{\:\prime \:2})}{{\vec{k}}_{{}}^{2}}\ln \left( \frac{\vec{q} 
_{1}^{\:2}}{\vec{q}_{2}^{\:2}}\right) +\;\;(\vec{q}_{1}\longleftrightarrow 
\vec{q}_{1}^{\:\prime },\;\;\vec{q}_{2}\longleftrightarrow \vec{q} 
_{2}^{\:\prime })\right\} .
\end{equation}

The quark contribution to the ``real'' part of the kernel in the Pomeron
channel, according to Eqs.~(\ref{z4}) and (\ref{z24})-(\ref{z26}) and (\ref{z45}),
can be presented as 
\begin{equation}
{\Large {\cal K}_{r}^{\left( 1\right) Q}\left( \vec{q}_{1},\vec{q}_{2}; 
\vec{q}\,\right) =}2{\Large {\cal K}_{r}^{\left( 8\right) Q}\left( \vec{q} 
_{1},\vec{q}_{2};\vec{q}\,\right) -}\frac{{\Large g^{4}n_{f}}}{(2\pi )^{5}N}K_{1}
\left(\vec{q}_{1}-\frac{\vec{q}}{2},\vec{q}_{2}-\frac{\vec{q}}{2}\right)~.
\label{z45}
\end{equation}

Let us mention the properties of the kernel: 
\[
{\cal K}_{r}^{({\cal R}) }( 0,\vec{q}_{2};\vec{q}\,) = 
{\cal K}_{r}^{({\cal R}) }( \vec{q}_{1},0;\vec{q}\,)
={\cal K}_{r}^{({\cal R}) }( \vec{q},\vec{q}_{2};\vec{q} 
\,) ={\cal K}_{r}^{({\cal  R}) }( \vec{q}_{1},\vec{q};\vec{q}\,) =0 ~,
\]
which follow from the gauge invariance, and  
\[
{\cal K}_{r}^{({\cal R}) }( \vec{q}_{1},\vec{q}_{2};\vec{q} 
\,) ={\cal K}_{r}^{({\cal  R}) }( \vec{q}_{2},\vec{q}_{1};- 
\vec{q}\,) ={\cal K}_{r}^{({\cal  R}) }( \vec{q} 
_{1}^{\: \prime},\vec{q}_{2}^{\: \prime };-\vec{q}\,) ~,
\]
which follow from the symmetry of the imaginary part of the
Reggeon-Reggeon scattering amplitude~(\ref{z6}), entering the
expression~(\ref{z5}) for the ''real'' part of the kernel. Let us stress
here that the above properties, descending from very general arguments, 
are valid also for the gluon part of the kernel; we omitted indeed 
the superscript $Q$ in the above equations. 

Another important property of the kernel is its infrared finiteness at fixed
${\vec{k}=}\vec{q}_{1}-\vec{q}_{2}.$ The $1/\epsilon $ singularity in
Eq.~(\ref{z44}) is the ultraviolet one and disappears when we expand the
kernel in terms of the renormalized coupling. Indeed, in this case we have
to add to the r.h.s. of Eq.~(\ref{z44}) the piece coming from the
coupling constant renormalization (see Eqs.~(\ref{z4a}) and (\ref{z4b})) 
in the LLA kernel (\ref{z3}). For the expansion in terms of the 
renormalized coupling we obtain in the limit $\epsilon \rightarrow 0$
\[
{\Large {\cal K}_{r}^{\left( 8\right) Q}\left( \vec{q}_{1},\vec{q}_{2}; 
\vec{q}\,\right) }_{renorm}{\Large =\frac{\bar{g}_{\mu }^{4}\mu ^{-2\epsilon
}n_{f}}{\pi ^{1+\epsilon }\Gamma (1-\epsilon )N}}
\]
\[
\times \left\{ \frac{1}{\epsilon }\left( \frac{2[\Gamma (2+\epsilon )]^{2}}{ 
\Gamma (4+2\epsilon )}\left( {\Large \frac{{\vec{k}}^{2}}{\mu ^{2}}}\right)
^{\epsilon }-\frac{1}{3}\right) \left[ \frac{{\Large (\vec{q}_{1}^{\:2}\vec{q} 
_{2}^{\:\prime \:2}+\vec{q}_{2}^{\:2}\vec{q}_{1}^{\:\prime \:2})}}{{\vec{k}}^{2}}- 
{\Large \vec{q}^{\:2}}\right] \right. 
\]

\begin{equation}
\left. +\frac{1}{3}\left[ \vec{q}^{\:2}\ln \left( \frac{\vec{q}^{\:2}\vec k^2}
{\vec{q}_{1}^{\:2}\vec{q}_{2}^{\:2}}\right) -\frac{{\Large (\vec{q}_{1}^{\:2} 
\vec{q}_{2}^{\:\prime \:2}-\vec{q}_{2}^{\:2}\vec{q}_{1}^{\:\prime \:2})}}{{\vec{k} 
}^{2}}\ln \left( \frac{\vec{q}_{1}^{\:2}}{\vec{q}_{2}^{\:2}}\right) \right] \
+\;\;(\vec{q}_{1}\longleftrightarrow \vec{q}_{1}^{\:\prime },\;\;\vec{q} 
_{2}\longleftrightarrow \vec{q}_{2}^{\:\prime })\right\} .
\end{equation}
As it can be seen from this expression, the $1/\epsilon$ singularity at fixed ${ 
\vec{k}}^{2}$ disappear after expanding $\left( \frac{\vec k^{2}}{\mu ^2}\right) 
^{\epsilon }$ in powers of $\epsilon$. The expansion is not performed here, 
since in the integral over ${\vec{k}}$ of Eq.~(\ref{z1}) the region of small 
${\vec{k}}^{2}$, for which $\epsilon \ln \left( \frac{\vec k^2}{\mu ^2}
\right) \sim 1$, does contribute. This region contributes to the integral
with terms singular in $1/\epsilon$. For the vacuum channel, these terms 
cancel the singularity in the ``virtual'' contribution to the kernel, related 
to the gluon trajectory. The cancellation occurs quite similarly to the forward 
case, so it does not need a special 
treatment. For the octet case such cancellation is evidently absent because of the 
different coefficients~(\ref{z4}) between vacuum and gluon channels.
We observe, however, that the cancellation is recovered in the case 
of the colourless compound state of three Reggeized gluons, i.e. the Odderon.
In this case, indeed, the ``real'' part of the kernel involves the three 
combinations with a different pair of Reggeized gluons in the octet channel, while
the ``virtual'' part of the kernel involves three gluon trajectories. The
cancellation of the infrared singularities follows then quite simply from the 
singular part of Eq.~(\ref{z45}).

\section{The check of the ``bootstrap'' condition}

The ``bootstrap'' condition derived in Ref.~\cite{FF98} has the form 
\[
\frac{g^2Nt}{2\left( 2\pi \right) ^{D-1}}\int \frac{d^{D-2}q_1}{\vec
q_1^{\:2}\left( \vec q_1-\vec q\right) ^2}\int \frac{d^{D-2}q_2}{\vec
q_2^{\:2}\left( \vec q_2-\vec q\right) ^2}{\cal K}^{\left( 8\right)
}\left( \vec q_1,\vec q_2;\vec q\right)
\]
\begin{equation}
=\omega ^{\left( 1\right) }\left( t\right) \left( \omega ^{\left( 1\right)
}\left( t\right) +\omega ^{\left( 2\right) }\left( t\right) \right) ~.
\label{z46}
\end{equation}
Here ${\cal K}^{\left( 8\right) }\left( \vec q_1,\vec q_2;\vec q\right) $
is the kernel of the non-forward BFKL equation, $\omega \left( t\right)
=\omega ^{\left( 1\right) }\left( t\right) +\omega ^{\left( 2\right) }\left(
t\right) $ is the deviation of the gluon Regge trajectory from unity in the
two-loop approximation and $t=-\vec q_{}^{\:2}.$ \quad In the one-loop
approximation (LLA) the trajectory 
\begin{equation}
\omega ^{\left( 1\right) }\left( t\right) =\frac{g^2Nt}{2\left( 2\pi \right)
^{D-1}}\int \frac{d^{D-2}q_1}{\vec q_1^{\:2}\left( \vec q_1-\vec q\right) ^2}
\label{z47}
\end{equation}
is purely gluonic. The quark contribution to the trajectory appears at the
two-loop level (NLLA) and is given by Eqs.~(\ref{z10}) and (\ref
{z11}). The kernel ${\cal K}^{\left( 8\right) }\left( \vec q_1,\vec
q_2;\vec q\right) ,$ according to Eq.~(\ref{z2}), is expressed through the
trajectory and the ``real'' part. The quark piece of the latter is given by
Eq.~(\ref{z44}). Using this equation together with Eqs.~(\ref{z10}) and (\ref
{z11}) for the quark contribution to the trajectory, we arrive at
\[
\int \frac{d^{D-2}q_2}{\vec q_2^{\:2}\left( \vec q_2-\vec q\right) ^2} 
{\cal K} ^{\left( 8\right) Q}\left( \vec q_1,\vec q_2;\vec q\right) = 
{\Large g^4n_fN\frac 1{(2\pi )^{D-1}}\frac 1{(4\pi )^{2+\epsilon }}\frac{ 
\Gamma (1-\epsilon )}\epsilon \frac{[\Gamma (2+\epsilon )]^2}{\Gamma
(4+2\epsilon )} }
\]
\[
\times \int \frac{d^{D-2}q_2}{\vec q_2^{\:2}}\left\{ {\Large \frac{\vec
q_1^{\:2}}{{\vec k}_{}^2}}\left( (\vec q_1^{\:2})^\epsilon +(\vec
q_1^{\:\prime \:2})^\epsilon -(\vec q_2^{\:2})^\epsilon -(\vec q_2^{\:\prime
\:2})^\epsilon \right) \right.
\]
\begin{equation}
\left. \ +\frac{{\Large \vec q^{\:2}}}{\vec q_2^{\:\prime \:2}}\left( ( 
{\Large \vec q^{\:2})}^\epsilon -(\vec q_1^{\:2})^\epsilon -(\vec
q_2^{\:2})^\epsilon \right) +\;\;(\vec q_1\longleftrightarrow \vec
q_1^{\:\prime })\right\} ~,  \label{z48}
\end{equation}
where ${\vec k=}\vec q_1-\vec q_2\:$, $\vec q_1^{\:\prime }=\vec q_1-\vec q$
and $\vec q_2^{\:\prime }=\vec q_2-\vec q$. Putting the r.h.s. expression
into Eq.~(\ref{z46}) and using Eqs.~(\ref{z47}), (\ref{z10}) and (\ref{z11}), it
is easy to verify that the ``bootstrap'' equation (\ref{z46}) is satisfied.

\section{Discussion}

In this paper we have calculated the quark part of the kernel of the
generalized non-forward BFKL equation at non-zero momentum transfer $t$ in
the next-to-leading logarithmic approximation. Along with the quark
contribution to the gluon Regge trajectory, which is the same as for the
case of zero momentum transfer $t$ and therefore is known already, this part
includes pieces coming from the quark contribution to the radiative corrections
for the one-gluon production and from the quark-antiquark production in the
Reggeon-Reggeon collisions. The results obtained can be used for an
arbitrary representation of the colour group in the $t-$channel. 

For all such representations, the part of the kernel related with
the real particle production is infrared finite, in the sense that
there are no singularities at fixed transverse momentum ${\vec{k}}$ of
the produced particles. The integration over ${\vec{k}}$ in the generalized
BFKL equation leads to terms singular in the limit $\epsilon \rightarrow 0$. 
For the vacuum channel, these terms cancel the singularity in the
''virtual'' contribution related to the gluon trajectory.  For the octet
case such cancellation is evidently absent, although it is recovered
in the case of the colourless compound state of three Reggeized gluons, i.e. for 
the Odderon.

The kernel for the octet case enters the ``bootstrap'' equation for
the gluon Reggeization in QCD. The fulfillment of this equation is necessary
for the self-consistency of the derivation of the BFKL equation. We
demonstrate explicitly the fulfillment of the ``bootstrap'' condition in the
next-to-leading logarithmic approximation in the part concerning the quark
contribution. The check performed serves simultaneously as a stringent
examination of the correctness of the calculations of the quark contribution to
the kernel of the BFKL equation. 

\bigskip

Recently a paper by M.~Braun and G.P.~Vacca~\cite{vacca} appeared, devoted to the 
NLLA kernel of the non-forward BFKL equation in the octet case. In this 
paper the kernel was obtained using as a basis the bootstrap relation and a 
specific ansatz to solve it. In Ref.~\cite{vacca} the kernel was defined as the 
kernel of the equation for the amputated Reggeized gluon - target amplitude 
(see Eqs.~(1)-(3) of Ref.~\cite{bra95} and Eq.~(2) of Ref.~\cite{vacca}), while 
our kernel is the kernel for the non-amputated amplitude (see our Eq.~(1)). 
This implies that the relation between our kernel, ${\cal K}_r$, 
and their kernel, $V$, should be (in the denotations used in Ref~\cite{vacca}) 
${\cal K}_r(q, q_1, q_1^{\: \prime}) = q_1^{\: \prime \: 2} q_2^{\: \prime \: 2}\: 
V(q, q_1, q_1^{\: \prime})$. 
With this correspondence our results disagree with those obtained 
in Ref.~\cite{vacca}. 

However, in a revised version of their paper, appeared after the present 
work was published as a preprint, the authors of Ref.~\cite{vacca} have 
added an Appendix where they use a different relation 
between their and our kernel (see Eq.~(51) of Ref.~\cite{vacca}). This relation 
follows from requiring that their ansatz for the kernel fulfills the correct 
symmetry properties. With this new correspondence there is indeed agreement 
between our results and theirs in the case of the quark contribution to the 
kernel. It would be interesting to check if the symmetrized ansatz leads 
to the correct results also for the gluon contribution to the kernel. 

\appendix

\section{Appendix I}

In this Appendix we present the details of the calculation of 
\begin{equation}
\int \frac{dk^2}{(2\pi )}d\rho _f A = \int_0^1\frac{dx}{2x(1-x)} \: \int 
\frac{d^{D-2}l_1}{(2\pi )^{(D-1)}} A ~,  \label{a1}
\end{equation}
where $A$ is given by Eq.~(\ref{z27}). We group the terms contributing to $A$
in four different classes according to their behaviour under the
integration. The first class contains only the first term in the r.h.s. of
Eq.~(\ref{z27}): 
\begin{equation}
A_1 = 32\frac{s\alpha _1\beta _2}{\tilde t_1\tilde t_1^{\prime }}(\vec
Q_1\vec Q_1^{\:\prime }) ~.
\end{equation}
It can be rewritten using Eqs.~(\ref{z17}), (\ref{z29}), (\ref{z30}) and ( 
\ref{z32}) in the equivalent form 
\begin{equation}
A_1 = 32 \frac{1-x}{x} \frac{\vec l_1 ^{\:2} (\vec Q_1\vec Q_1^{\:\prime})} { 
\tilde t_1 \tilde t_1^{\prime}} = \frac{ 32x(1-x) \, \vec l_1 ^{\:2} (\vec
Q_1\vec Q_1^{\:\prime})}{[\vec l_1 ^{\:2}+x(\vec q_1^{\:2} - 2 (\vec l_1
\vec q_1))]\: [\vec l_1 ^{\:2}+x(\vec q_1 ^{\:\prime \:2} - 2 (\vec l_1 \vec
q_1 ^{\: \prime}))] } ~.
\end{equation}
Taking first the integral over $x$, we have 
\[
\int_0^1\frac{dx}{2x(1-x)} \: A_1 = \frac{ 16 (\vec Q_1\vec Q_1^{\:\prime}) 
} {(\vec q_1^{\:\prime} - \vec l_1)^2 - (\vec q_1 - \vec l_1)^2 } \ln\frac{ 
(\vec q_1^{\:\prime} - \vec l_1)^2} {(\vec q_1 - \vec l_1)^2} ~.
\]
Using now the following simple trick 
\[
\frac{1} {(\vec q_1^{\:\prime} - \vec l_1)^2 - (\vec q_1 - \vec l_1)^2 } \ln 
\frac{(\vec q_1^{\:\prime} - \vec l_1)^2} {(\vec q_1 - \vec l_1)^2} =
\int_0^1 dz \: \frac{1}{z(\vec q_1^{\:\prime} - \vec l_1)^2 + (1-z)(\vec
q_1-\vec l_1)^2 } ~,
\]
the integration over $l_1$ and the subsequent integration over $z$ become
trivial and give 
\begin{equation}
- \frac{64 \Gamma (-\epsilon)}{(4\pi )^{2+\epsilon }} \frac{[\Gamma
(2+\epsilon )]^2}{\Gamma (4+2\epsilon )} (\vec q^{\:2})^{1+\epsilon } + 16
\int \frac{d^{D-2}l_1}{(2\pi )^{(D-1)}} ~.
\end{equation}
The last term in the above expression vanishes in dimensional
regularization. We stress once more, however, that independently from the
regularization scheme, the integrals which diverge at large $|\vec l_i|$ in $ 
D=4$ cancel in Eq.~(\ref{a1}).

The second class of terms in the r.h.s. of Eq.~(\ref{z27}) is 
\[
A_2 = 8-\frac{16}{k^2}(\vec q^{\:2}-\frac{\vec q_1^{\:2}\vec q_2^{\:\prime
\:2}+\vec q_2^{\:2}\vec q_1^{\:\prime \:2}}{s\alpha \beta })
\]
\begin{equation}
-\frac 8{(k^2)^2}[s\alpha \beta _1-s\alpha _1\beta +2\vec q_1^{\:2}\frac{
\alpha _1}\alpha -2\vec q_2^{\:2}\frac{\beta _1}\beta -2\vec l_1(\vec
q_1+\vec q_2)]
\end{equation}
\[
\times [s\alpha \beta _1-s\alpha _1\beta +2\vec q_1^{\:\prime \:2}\frac{ 
\alpha_1} \alpha -2\vec q_2^{\:\prime \:2}\frac{\beta _1}\beta -2\vec
l_1(\vec q_1^{\:\prime }+\vec q_2^{\:\prime })] ~.
\]
Using again Eqs.~(\ref{z17}), (\ref{z29}), (\ref{z30}) and (\ref{z32}), the
above expression can be recast in the following form: 
\[
A_2 = 16x(1-x)\left[ 2 + (1-2x)^2 \frac{\vec q_1^{\:2} + \vec q_1^{\:\prime
\:2}} {\Sigma} - 2x(1-x)(1-2x)^2 \frac{\vec q_1^{\:2}\vec q_1^{\:\prime \:2} 
}{\Sigma^2} \right]
\]
\[
-\frac{16x^2(1-x)^2}{\vec \Lambda^2}\left[ \frac{2(1-2x)}{x(1-x)}\left(
(\vec \Lambda \vec q_1^{\:\prime})-x(1-x)\frac{\vec q_1^{\:\prime \:2} (\vec
\Lambda \vec k)}{\Sigma} \right)\left(1-2x(1-x)\frac{\vec q_1^{\:2}}{\Sigma} 
\right)\right.
\]
\begin{equation}
+ \frac{2(1-2x)}{x(1-x)}\left( (\vec \Lambda \vec q_1)-x(1-x)\frac{\vec
q_1^{\:2} (\vec \Lambda \vec k)}{\Sigma} \right)\left(1-2x(1-x)\frac{\vec
q_1^{\:\prime \:2}}{\Sigma}\right)
\end{equation}
\[
\left. - \frac{\vec q_1^{\:\prime \:2} \vec q_2^{\:2} + \vec q_1^{\:2} \vec
q_2^{\:\prime \:2}} {\Sigma} + \frac{\vec q^{\:2}}{x(1-x)} \right]
\]
\[
- \frac{128x^2(1-x)^2}{(\vec \Lambda^2 )^2} \left((\vec \Lambda \vec
q_1)-x(1-x)\frac{\vec q_1^{\:2} (\vec \Lambda \vec k)}{\Sigma} \right)
\left( (\vec \Lambda \vec q_1^{\:\prime})-x(1-x)\frac{\vec q_1^{\:\prime
\:2} (\vec \Lambda \vec k)}{\Sigma} \right) ~.
\]
The non-zero integrals over $l_1$ are the following: 
\[
I_1 = \int \frac{d^{D-2}l_1}{(2\pi )^{(D-1)}} \frac{1}{\Sigma} = \frac{2
\Gamma (-\epsilon)} {(4\pi )^{2+\epsilon }} [x(1-x){\vec k}_{}^2]^\epsilon ,
\]

\begin{equation}
I_2 = \int \frac{d^{D-2}l_1}{(2\pi )^{(D-1)}} \frac{1}{\Sigma^2} = \frac{2
\Gamma (1-\epsilon)} {(4\pi )^{2+\epsilon }} [x(1-x){\vec k}_{}^2 
]^{\epsilon-1} ~,
\end{equation}

\[
I_3 = \int \frac{d^{D-2}l_1}{(2\pi )^{(D-1)}} \frac{1}{\vec \Lambda^2 \Sigma}
= -\frac{2 \Gamma (-\epsilon)} {(4\pi )^{2+\epsilon }} [x(1-x){\vec k}_{}^2 
]^{\epsilon-1} ~,
\]

\[
I_4 = \int \frac{d^{D-2}l_1}{(2\pi )^{(D-1)}} \frac{1} {(\vec \Lambda^2)^2} 
\left((\vec \Lambda \vec q_1)-x(1-x)\frac{\vec q_1^{\:2} (\vec \Lambda \vec
k)}{\Sigma} \right) \left( (\vec \Lambda \vec q_1^{\:\prime})-x(1-x)\frac{ 
\vec q_1^{\:\prime \:2} (\vec \Lambda \vec k)}{\Sigma} \right)
\]
\[
= - \frac{\Gamma (-\epsilon)} {(4\pi )^{2+\epsilon }} \frac{(1-\epsilon)\vec
q_1^{\:2}\vec q_1^{\:\prime \:2} - \vec q_1^{\:\prime \:2} (\vec q_1 \vec k)
- \vec q_1^{\:2} (\vec q_1^{\:\prime} \vec k)}{(1+\epsilon)} x (1-x) [x(1-x){ 
\vec k}_{}^2]^{\epsilon-1} ~.
\]
Using the above result and integrating also over $x$ , we finally obtain 
\begin{equation}
\int_0^1\frac{dx}{2x(1-x)} \: \int \frac{d^{D-2}l_1}{(2\pi )^{(D-1)}} A_2 =
- \frac{64 \Gamma (-\epsilon)}{(4\pi )^{2+\epsilon }} \frac{[\Gamma
(2+\epsilon )]^2}{\Gamma (4+2\epsilon )} \frac{({\vec k}_{}^2)^\epsilon}{{ 
\vec k}_{}^2} (\vec q_1^{\:\prime\:2} \vec q_2^{\:2}+\vec q_1^{\:2} \vec
q_2^{\:\prime\:2}) ~.
\end{equation}

The remaining terms in the r.h.s. of Eq.~(\ref{z27}) are 
\[
A_{3}=\frac{16}{k^{2}}\left[ \frac{1}{\tilde{t}_{1}}\left( ((\vec{Q}_{1}\vec{ 
q}_{1}^{\:\prime })(s\alpha _{1}\beta _{2}-\vec{l}_{1}\vec{l}_{2})+(\vec{Q} 
_{1}\vec{l}_{2})(s\alpha _{1}\beta +\vec{l}_{1}(\vec{q}_{1}^{\:\prime }+\vec{q 
}_{2}^{\:\prime })-2\vec{q}_{1}^{\:\prime \:2}\frac{\alpha _{1}}{\alpha } 
))\right. \right. 
\]
\begin{equation}
\left. \left. +(\vec{l}_{1}\leftrightarrow \vec{l}_{2},\alpha
_{1}\leftrightarrow \beta _{2},\alpha _{2}\leftrightarrow \beta _{1},\vec{q} 
_{1}^{~}\leftrightarrow -\vec{q}_{2}^{~},\vec{q}_{1}^{\:\prime
}\leftrightarrow -\vec{q}_{2}^{\:\prime })\phantom{\frac{.}{.}}\!\!\right)  
\right] 
\end{equation}
and 
\begin{equation}
A_{4}=A_{3}(\,\vec{q}_{1}^{~}\leftrightarrow \vec{q}_{1}^{\:\prime },\quad 
\vec{q}_{2}^{~}\leftrightarrow \vec{q}_{2}^{\:\prime })~.
\end{equation}
It is not necessary to calculate explicitly the integral of $A_{4}$, since
it can be obtained by simple substitutions from that of $A_{3}$. Concerning $ 
A_{3}$, it can be rewritten equivalently as 
\[
A_{3}=\frac{16x(1-x)}{\vec{\Lambda}^{2}\tilde{t}_{1}}\left[ (\vec{l}_{1}\vec{ 
Q}_{1})\,\vec{l}_{2}(\vec{q}_{1}^{\:\prime }+\vec{q}_{2}^{\:\prime })+(\vec{l} 
_{2}\vec{Q}_{1})\,\vec{l}_{1}(\vec{q}_{1}^{\:\prime }+\vec{q}_{2}^{\:\prime })+ 
\frac{(\vec{\Lambda}\vec{l}_{1})\,\vec{Q}_{1}(\vec{q}_{1}^{\:\prime }+\vec{q} 
_{2}^{\:\prime })}{x}\right. 
\]
\begin{equation}
\left. +\alpha _{1}\beta s\left( 1-\frac{2\vec{q}_{1}^{\:\prime \:2}}{\alpha
\beta s}\right) (\vec{l}_{2}\vec{Q}_{1})-\alpha \beta _{2}s\left( 1-\frac{2 
\vec{q}_{2}^{\:\prime \:2}}{\alpha \beta s}\right) (\vec{l}_{1}\vec{Q}_{1}) 
\right] ~.  \label{a2}
\end{equation}
Since $A_{3}$ is manifestly invariant under the ``left-right''
transformation~(\ref{z37}), we can separate in the above expression two set
of terms, related each other by the ``left-right'' transformation. One
possible separation is 
\begin{eqnarray}
A_{3} &=&\frac{16x(1-x)}{\vec{\Lambda}^{2}\tilde{t}_{1}}\left[ (\vec{l}_{1} 
\vec{Q}_{1})\,\vec{l}_{2}(\vec{q}_{1}^{\:\prime }+\vec{q}_{2}^{\:\prime })+ 
\frac{(\vec{\Lambda}\vec{l}_{1})\,\vec{Q}_{1}(\vec{q}_{1}^{\:\prime }+\vec{q} 
_{2}^{\:\prime })}{2x}-\alpha \beta _{2}s\left( 1-\frac{2\vec{q}_{2}^{\:\prime
\:2}}{\alpha \beta s}\right) (\vec{l}_{1}\vec{Q}_{1})\right]   \nonumber \\
&+&\left[ \mbox{``left-right''}\phantom{\frac{.}{.}}\!\!\right] \quad \equiv
\quad f_{3}+f_{3}^{(L/R)}~,
\end{eqnarray}
with obvious notation. Since the integration measure can be put in the two
equivalent forms~(\ref{z35}) and (\ref{z38}) connected by the ``left-right''
transformation, the result of the integration of $f_{3}^{(L/R)}$ can be
obtained from that of $f_{3}$ by the change $(\,\vec{q}_{1}^{~} 
\leftrightarrow -\vec{q}_{2}^{~},\quad \vec{q}_{1}^{\:\prime }\leftrightarrow
-\vec{q}_{2}^{\:\prime })$. Therefore the integration of $f_{3}^{(L/R)}$
can be avoided. This allows to escape those integrands with 
$\vec{\Lambda}^{2}\,\tilde{t} 
_{1}\Sigma $ at the denominator which come from the term proportional to $ 
\alpha _{1}/\alpha $ in Eq.~(\ref{a2}) and would be very nasty to integrate
with the measure~(\ref{z35}). Let us focus then our attention on $f_{3}$
which can be written as
\begin{eqnarray}
f_{3} &=&-\frac{8x(1-x)}{\vec{\Lambda}^{2}\tilde{t}_{1}}\left\{ \frac{2}{x} 
\vec{\Lambda}^{2}(\vec{l}_{1}\vec{Q}_{1})-\frac{(\vec{\Lambda}\vec{l}_{1})\, 
\vec{Q}_{1}(\vec{q}_{1}^{\:\prime }+\vec{q}_{2}^{\:\prime })}{x}\right.  
\nonumber \\
&+&\left. 2(\vec{l}_{1}\vec{Q}_{1})[-2(1-x)(\vec{q}_{1}^{\:\prime }\vec{q} 
_{2}^{\:\prime })+x(\vec{q}_{2}^{\:\prime \:2}-\vec{q}_{1}^{\:\prime \:2})+\vec{l} 
_{1}(\vec{q}_{1}^{\:\prime }+\vec{q}_{2}^{\:\prime })]\phantom{\frac{.}{.}} 
\!\!\right\} ~.  \label{a3}
\end{eqnarray}
The integration of the first term is trivial and gives 
\begin{equation}
I_{5}=-\int_{0}^{1}\frac{dx}{2x(1-x)}\int \frac{d^{D-2}l_{1}}{(2\pi )^{(D-1)} 
}\left[ \frac{16(1-x)\vec{\Lambda}^{2}(\vec{l}_{1}\vec{Q}_{1})}{\vec{\Lambda} 
^{2}\tilde{t}_{1}}\right] =\frac{32\Gamma (-\epsilon )}{(4\pi )^{2+\epsilon } 
}\frac{[\Gamma (2+\epsilon )]^{2}}{\Gamma (4+2\epsilon )}(\vec{q} 
_{1}^{\:2})^{1+\epsilon }~.
\end{equation}
For the remaining terms we limit ourselves to illustrate the strategy of the
integration, since presenting all the intermediate results would be too
lengthy. The basic integrals to be calculated are of the form\footnote{ 
Strictly speaking, there are also integrals with $(\vec{l}_{1}\vec{p})$, $ 
\vec{l}_{1}^{\;2}$ or $(\vec{l}_{1}\vec{p})\vec{l}_{1}^{\;2}$ at the
numerator, where $\vec{p}$ is a generic momentum in the transverse space,
but they can be treated similarly.} 
\begin{equation}
I=\int_{0}^{1}dx\int \frac{d^{D-2}l_{1}}{(2\pi )^{(D-1)}}\frac{x^{n+1}}{( 
\vec{l}_{1}-x\vec{k})^{2}\,[(\vec{l}_{1}-x\vec{q}_{1})^{2}+x(1-x)\vec{q} 
_{1}^{\:2}]}~,
\end{equation}
with $n$ natural number. Using the Feynman parametrization and integrating
in $d^{D-2}l_{1}$, one obtains 
\[
I=\frac{2\Gamma (1-\epsilon )}{(4\pi )^{2+\epsilon }}\int_{0}^{1}dx 
\int_{0}^{1}dz\frac{x^{n+1}}{\{xz[x(1-z)\vec{q}_{2}^{\:2}+(1-x)\vec{q} 
_{1}^{\:2}]\}^{1-\epsilon }}
\]
\begin{equation}
=\frac{2\Gamma (1-\epsilon )}{(4\pi )^{2+\epsilon }}\int_{0}^{1}dy\,y^{ 
\epsilon -1}\int_{y}^{1}dx\frac{x^{n}}{[x(\vec{q}_{2}^{\:2}-\vec{q}_{1}^{\:2})+ 
\vec{q}_{1}^{\:2}-y\vec{q}_{2}^{\:2}]^{1-\epsilon }}~,
\end{equation}
where the change of variable $y=xz$ has been performed in the last equality.
This integral can be now calculated integrating first over $x$ and then over 
$y$. The complete calculation for all the terms in Eq.~(\ref{a3}) except the
first is long, but straightforward. The final result for 
\begin{equation}
\int_{0}^{1}\frac{dx}{2x(1-x)}\:\int \frac{d^{D-2}l_{1}}{(2\pi )^{(D-1)}} 
(A_{3}+A_{4})
\end{equation}
is given by the last three rows in the r.h.s. of Eq.~(\ref{z40}).

\end{document}